\documentclass[twocolumn,letterpaper,aps,prl,
superscriptaddress,showpacs,amsmath]{revtex4}

\usepackage{graphicx}

\newcommand{\vc}[1]{\boldsymbol{#1}}
\newcommand{\im}{\mathrm{i}}

\begin{document}

\title{Spin polaron theory for the photoemission spectra of layered cobaltates}

\author{Ji\v{r}\'{\i} Chaloupka}
\affiliation{Max-Planck-Institut f\"ur Festk\"orperforschung,
Heisenbergstrasse 1, D-70569 Stuttgart, Germany}
\affiliation{Department of Condensed Matter Physics, Faculty of Science, 
Masaryk University, Kotl\'a\v{r}sk\'a 2, 61137 Brno, Czech Republic}

\author{Giniyat Khaliullin}
\affiliation{Max-Planck-Institut f\"ur Festk\"orperforschung,
Heisenbergstrasse 1, D-70569 Stuttgart, Germany}

\begin{abstract}
Recently, strong reduction of the quasiparticle peaks and pronounced 
incoherent structures have been observed in the photoemission spectra of 
layered cobaltates. Surprisingly, these many-body effects are found to 
increase near the band insulator regime. We explain these unexpected 
observations in terms of a novel spin-polaron model for CoO$_2$ planes 
which is based on a fact of the spin-state quasidegeneracy of Co$^{3+}$ ions 
in oxides. Scattering of the photoholes on spin-state fluctuations suppresses 
their coherent motion. The observed ``peak-dip-hump'' type lineshapes are well 
reproduced by the theory. 
\end{abstract}

\date{\today}



\maketitle

Strongly correlated behavior of electrons is a common property of transition
metal oxides. This is because the bandwidth in these compounds is relatively
small compared to the intraionic Coulomb repulsion between the $3d$ electrons.
As a result, the celebrated Mott physics \cite{Mot74} forms a basis for
understanding of a unique properties of oxides, such as the high-$T_c$
superconductivity and a colossal magnetoresistivity.   

Recently, attention has focused on the layered cobalt oxides because they
exhibit  high thermoelectric power \cite{Wan03}, i.e., the capability to
transform heat energy into electricity. These compounds consist of a
triangular lattice CoO$_2$ planes, separated either by Na layers like in
NaCoO$_2$ \cite{Foo04} or by thick layers of rock-salt structure in so-called
``misfit'' cobaltates (see \cite{Bob07,Bro07} and references therein).
Besides controlling the $c$-axis transport, the Na and rock-salt layers
introduce also the charge carriers into the CoO$_2$ planes, such that the
valence state of Co ions is varied in a wide range from nonmagnetic
Co$^{3+}t_{2g}^6$ $S=0$ state (as in NaCoO$_2$) towards the magnetic
Co$^{4+}t_{2g}^5$ $S=1/2$ configuration (in Na$_x$CoO$_2$ at small $x$).  

As the $t_{2g}^6$ shell of Co$^{3+}$ is full, this limit is naturally referred
to as a band insulator \cite{Lan05,Vau05}, while Co$^{4+}$ $S=1/2$ rich
compounds fall into the category of Mott systems because of unquenched spins.
It has therefore been thought that layered cobaltates may provide an
interesting opportunity to monitor the evolution of electronic states from a
weakly correlated band-insulator regime to the strongly correlated Mott-limit
by hole doping of NaCoO$_2$. Surprisingly, a completely opposite trend is
found experimentally. The hallmarks of strong correlations such as magnetic
order \cite{Foo04,Bay05}, strong magnetic field effects \cite{Wan03}, {\it
etc,}  are most pronounced closer to the Co$^{3+}$ compositions, while
Co$^{4+}$ $S=1/2$ rich compounds behave as a moderately correlated metals
\cite{Foo04,Vau07}.  The best thermoelectric performance of cobaltates is also
realized near the doped band insulator regime \cite{Lee06}, thus unusual
correlations in this state and enhanced thermopower are clearly interrelated.
As a direct evidence of a complex structure of doped holes, the recent angular
resolved photoemission (ARPES) experiments \cite{Qia06,Bro07} observed
lineshapes that are typical for strongly correlated systems.  Paradoxically
again, the many-body effects in ARPES are enhanced approaching the band
insulator limit~\cite{Bro07}. 

This Letter presents a theory resolving this puzzling situation in layered
cobaltates. We show that holes doped into nonmagnetic band insulators
NaCoO$_2$ and misfits are indeed a composite objects with a broad
energy-momentum distribution of their spectral functions. Besides a reduction
of the quasiparticle peaks, they display also a dispersive incoherent
structure as observed \cite{Qia06,Bro07}. The underlying physics behind this
unexpected complexity is in fact quite simple and based on a unique aspect of
Co$^{3+}$ ions, i.e., their spin-state quasidegeneracy, and on special lattice
geometry of CoO$_2$ layers. 

In oxides, Co$^{2+}$ is always in a high-spin $3/2$ state while Co$^{4+}$ ions
usually adopt a low-spin $1/2$. Roughly, this selection is decided by the Hund
coupling favoring high-spin values, or by the  $t_{2g}$--$e_g$ crystal field
splitting $10Dq$ supporting low-spin states.  An intermediate case is realized
for the Co$^{3+}$ ions where the $S=0$,~$1$,~$2$ states strongly compete. This
leads to a distinct property Co$^{3+}$ rich oxides called
``spin-state-transition'', which is responsible for many anomalies such as
thermally or doping induced spin-state change in LaCoO$_3$ \cite{Yam96,Hav06},
and  a spin-blockade effect in HoBaCo$_2$O$_{5.5}$ \cite{Mai04}, to mention a
few manifestations of the Janus-like behavior of Co$^{3+}$.  

\begin{figure}[tbp]
\begin{center}
\includegraphics[width=7.5cm]{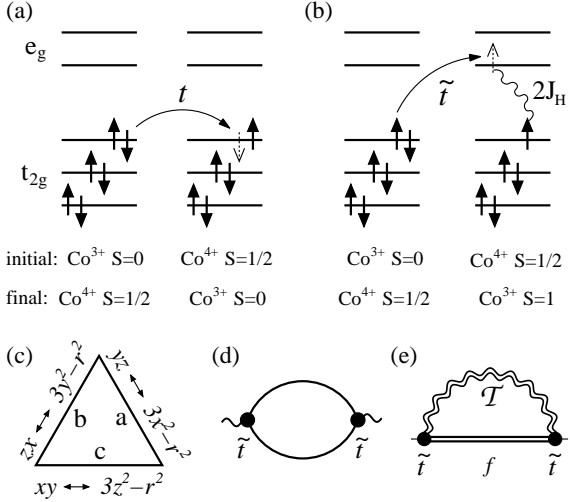}
\caption{
(a) A conventional $t$-hopping within the low-spin $t_{2g}$ states. 
(b) $\tilde{t}$-hopping process creating an excited Co$^{3+}$ $S=1$ state.
Because of the Hund coupling $J_H$, this triplet state is quasidegenerate 
to the $t_{2g}^6$~$S=0$ state, i.e. the excitation energy $E_T\sim 10Dq-2J_H$ 
is low. 
(c) Bond directions $a$, $b$, $c$ in the triangular lattice of Co ions and 
$\tilde{t}$-coupled orbitals on these bonds.
(d) Selfenergy of the ${\cal T}$ spin excitation.
(e) Selfenergy of the holes in a selfconsistent Born approximation.
}
\label{fig:schemas}
\end{center}
\end{figure}

New situation encountered in layered cobaltates -- and this is the key point
-- is that CoO$_6$ octahedra are edge-shared. In this geometry, the charge
transfer between Co ions occurs along the 90$^{\circ}$ $d$-$p$-$d$ bonds,
where the largest matrix element for the hole motion is that between the
neighboring orbitals of $t_{2g}$ and $e_g$ symmetry (see $\tilde t$ process in
Fig.~\ref{fig:schemas}). This implies that a doped hole dynamically creates
$S=1$ $t_{2g}^5e_g^1$ states of Co$^{3+}$ \cite{note1}.  It is this point
where the holes -- doped into an initially nonmagnetic background -- become
many-body objects dressed by virtual spin excitations.  Formation of spin
polarons suppresses the plane-wave like motion of holes.  However, they gain
in the kinetic energy by exploiting an additional, $\tilde t$ hopping channel.           
  
The layered structure of cobaltates results in quasi two-dimensional
electronic states like in cuprates. This should also lower the symmetry of the
$t_{2g}$ bands (threefold-degenerate in the case of a cubic lattice) that
accommodate doped holes. Indeed, ARPES data shows a very simple Fermi surface
derived from a single band of $a_{1g}$ symmetry \cite{Qia06,Bro07}.  This band
can be described by the nearest-neighbor Hamiltonian
$H_t=-t\sum_{ij\sigma} f^\dagger_{j\sigma}f^{\phantom{\dagger}}_{i\sigma}$, 
where the fermionic operators $f$ represent $a_{1g}$ holes. They are subject
to a conventional Gutzwiller constraint (at most one hole at given site) but
this is not much relevant at small density of holes, i.e., near the
band-insulator limit. Hence, holes move freely as in a semiconductor.  

In contrast, the $t_{2g}$--$e_g$ hopping in Fig.~\ref{fig:schemas}(b)
represents the many-body process as it produces the Co$^{3+}$ $S=1$ excitation. 
The model describing this process has been derived in Ref. \cite{Kha07} 
and reads as follows:   
\begin{multline}
\label{Htilde}
H_{\tilde t} = -\frac{\tilde t}{\sqrt 3} \sum_{ij} \Bigl[
{\cal T}^\dagger_{+1,\gamma}(i)
f^\dagger_{j\downarrow}f^{\phantom{\dagger}}_{i\uparrow}-
{\cal T}^\dagger_{-1,\gamma}(i)
f^\dagger_{j\uparrow}f^{\phantom{\dagger}}_{i\downarrow}
\\
-{\cal T}^\dagger_{0,\gamma}(i)\,\tfrac1{\sqrt{2}}\!\left(
  f^\dagger_{j\uparrow}f^{\phantom{\dagger}}_{i\uparrow}-
  f^\dagger_{j\downarrow}f^{\phantom{\dagger}}_{i\downarrow}
\right)
+\mathrm{h.c.}\Bigr].
\end{multline} 
Here, ${\cal T}$ is the spin-triplet excitation generated by hopping of an
electron from Co$^{3+}_j$ to the $e_g$ level of Co$^{4+}_i$ (described as hole
motion).  The process, of course, conserves the total spin.  In addition, the
direction $\gamma$ of the bond $\langle ij \rangle$ selects the $e_g$ orbital
involved in $\tilde{t}$ process [see Fig.\ref{fig:schemas}(c)], according to
$E_g$ symmetry relations between ${\cal T}_{\gamma}$ operators:   
${\cal T}_c={\cal T}_{3z^2-r^2}$, 
${\cal T}_{a/b}=-\frac{1}{2}{\cal T}_{3z^2-r^2} 
\pm \frac{\sqrt 3}{2}{\cal T}_{x^2-y^2}$. 

Based on this model, we develop a theory for the photoemission experiments in
cobaltates. It is evident from \eqref{Htilde}, that by creating and destroying
${\cal T}$ excitations as they propagate, the holes are strongly renormalized
and we deal with a spin-polaron problem.  This resembles the problem of doped
Mott insulators like cuprates, however, the nature of spin excitations is
different here because of the nonmagnetic ground state.  Instead of
magnon-like propagating modes as in cuprates, fluctuations of the very spin
value of Co$^{3+}$ ions are the cause of the spin-polaron physics in
cobaltates \cite{note2}.   

For the calculation of the fermionic self-energies, we employ the
selfconsistent Born approximation [see Fig.~\ref{fig:schemas}(e)], which has
extensively been used in the context of spin-polarons in cuprates
\cite{Kan89}. First, we focus on spin excitation spectrum.  Since a direct
$e_g$-$e_g$ hopping in case of $90^{\circ}$-bonds is not allowed by symmetry,
the bare ${\cal T}$ spin excitation is a purely local mode,  at the energy
$E_T$. The coupling to the holes in \eqref{Htilde} shifts and broadens this
level. Accounting for this effect perturbatively [see
Fig.~\ref{fig:schemas}(d)], we obtain the ${\cal T}$ Green's function 
${\cal D}^{-1}(\im\omega)=\im\omega-E_T-\Sigma_T(\im\omega)$ with 
\begin{equation}
\label{tripselfE}
\Sigma_T(\im\omega)=\frac{2\tilde{t}^{\,2}}{3\beta}
\sum_{\vc k\vc k',\im\epsilon} \Gamma_{\vc k}\,
{\cal G}_0(\vc k,\im\epsilon) {\cal G}_0(\vc k',\im\epsilon+\im\omega) \;.
\end{equation}
Here, ${\cal G}_0$ is the bare electron propagator
${\cal G}_0(\vc k,\im\epsilon)=(\im\epsilon -\xi_{\vc k})^{-1}$
with the $a_{1g}$ dispersion on a triangular lattice 
$\xi_{\vc k}=-2t(c_a+c_b+c_c)+\mu$, 
where $c_\gamma=\cos k_\gamma$ and  $k_\gamma$ are the projections of $\vc k$
on $a$,$b$,$c$ directions.  The underlying $E_g$ symmetry of ${\cal T}$
operators involved in $\tilde{t}$ hopping results in the factor 
$\Gamma_{\vc k}=c_a^2+c_b^2+c_c^2-c_ac_b-c_bc_c-c_cc_a$. 
We neglected a weak momentum dependence of $\Sigma_T$ for the sake of
simplicity. This is justified as long as $\Sigma_T$ is small compared to the
spin gap $E_T$. 

Further, we approximate $\Gamma_{\vc k}$ by its Brillioun-zone average $3/2$,
obtaining the simple expressions for $\Sigma_T$ in terms of bare fermionic
density of states $N_0(x)=\sum_{\vc k} \delta(x-\xi_{\vc k})$: 
\begin{equation}
\label{ImET}
\mathrm{Im}\Sigma_T(E)=
 -\pi\tilde{t}^{\,2}\int_{-E}^0 \mathrm{d}x\, N_0(x)N_0(x+E) \;,
\end{equation}
\begin{equation}
\label{ReET}
\mathrm{Re}\Sigma_T(E)=
 -\tilde{t}^{\,2}\int_{-\infty}^0\mathrm{d}x
\int_{x^2}^\infty \mathrm{d}y^2\,
\frac{N_0(x)N_0(x+y)}{y^2-E^2} \;, 
\end{equation}
These equations determine the renormalized spin-excitation spectrum 
$\rho_T(E)=-\pi^{-1}\mathrm{Im}{\cal D}(\im\omega \rightarrow E+\im\delta)$
used below for calculation of the fermionic selfenergy. 

The selfenergy diagram in Fig.~\ref{fig:schemas}(e) reads as:  
\begin{equation}\label{holeselfE}
\Sigma_{\vc k}(\im\epsilon)=-\frac{2\tilde{t}^{\,2}}{\beta}
\sum_{\vc k', \im\omega}\left[
\Gamma_{\vc k'} {\cal D}(-\im\omega)+
\Gamma_{\vc k} {\cal D}(\im\omega)\right]
{\cal G}(\vc k',\im\epsilon+\im\omega) \;,
\end{equation}
where ${\cal G}^{-1}(\vc k,\im\epsilon)= 
\im\epsilon-\xi_{\vc k}-\Sigma_{\vc k}(\im\epsilon)$. 
We can write 
$\Sigma_{\vc k}(\omega)=
2\tilde{t}^{\,2}[\Phi(\omega)+\Gamma_{\vc k}\Xi(\omega)]$, 
where 
\begin{equation}\label{Phi}
\Phi(\omega)=\int_0^\infty\mathrm{d}E\, \rho_T(E)
\int_0^\infty \mathrm{d}x\, \frac{\tilde{N}(x)}{\omega-E-x+\im\delta} \;,
\end{equation}
\begin{equation}\label{Xi}
\Xi(\omega)=\int_0^\infty\mathrm{d}E\, \rho_T(E)
\int_{-\infty}^0 \mathrm{d}x\, \frac{N(x)}{\omega+E-x+\im\delta} \;.
\end{equation}
The full local density of states $N(E)=\sum_{\vc k} A({\vc k},E)$ and 
its $E_g$ symmetry part 
$\tilde{N}(E)=\sum_{\vc k} \Gamma_{\vc k} A({\vc k},E)$ are functions of
the selfenergy itself, via the spectral functions 
$A({\vc k},E)=-\pi^{-1}\mathrm{Im}{\cal G}(\im\epsilon\rightarrow
E+\im\delta)$. 
The above equations are thus to be solved self-consistently. 

\begin{figure}[tbp]
\begin{center}
\includegraphics{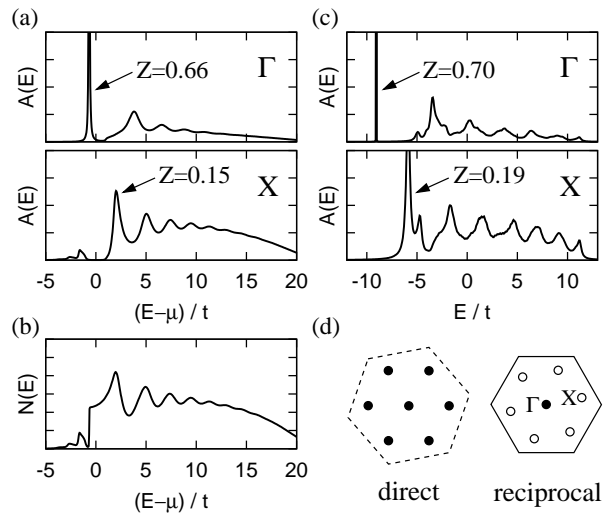}
\caption{
Comparison of results from the diagrammatic calculation at $10\:\%$ doping
(left) and from exact diagonalization (right). 
(a) Spectral functions and a quasiparticle weight $Z$ obtained from
Eqs.~\eqref{ImET}--\eqref{Xi} with bare $E_T=1.2t$ (renormalized to
$\tilde{E}_T/t\approx1$ by interactions) and
(b) the corresponding density of states $N(E)$. The incoherent structure
dominates $N(E)$. A peak around $2t$ corresponds to the van Hove singularity
smeared by interaction effects. 
(c) Spectral functions from exact diagonalization at $E_T/t=1$.  
(d) The 7-site cluster in direct space (the dashed line defines the supercell)
and in reciprocal space (full line is the Brillouin zone boundary). The
allowed $\vc k=\Gamma,X$ points are indicated by ($\bullet$) and ($\circ$). 
}
\label{fig:profiles}
\end{center}
\end{figure}

Next, we would like to test the reliability of approximations made.  To this
end, we performed an exact diagonalization of our model $H=H_t+H_{\tilde{t}}$
on a $\sqrt{7}\times\sqrt{7}$ hexagonal cluster [Fig.~\ref{fig:profiles}(d)].
We inject one hole on such a cluster and calculate its spectral function 
\begin{equation}\label{AkEexact}
A({\vc k},E)=- \frac1\pi \mathrm{Im}
\langle GS| f_{{\vc k},\sigma}(z-H)^{-1} 
f^\dagger_{{\vc k},\sigma} |GS\rangle \;,
\end{equation}
where $z=E+E_{GS}+\im\delta$.  Since the groundstate $|GS\rangle$ contains no
holes and the single hole on the cluster represents $1/7$ doping, the
effective doping is about 10\%. Periodic boundary conditions allow us to
access $A({\vc k},E)$ at two non-equivalent ${\vc k}$ points [see
Fig.~\ref{fig:profiles}(d)]. To obtain continuous profiles of $A({\vc k},E)$,
we have broadened the excited states. 

Our model has two parameters: $\tilde{t}/t$ and $E_T/t$. In fact, the ratio 
$\tilde{t}/t\simeq 3$ follows from the relations $t=2t_0/3$ and 
$\tilde{t}/t_0=t_{\sigma}/t_{\pi} \simeq 2$ (where
$t_0=t_{\pi}t_{\pi}/\Delta_{pd}$) \cite{Kha07}. Thus, we set 
below $\tilde{t}/t=3$, leaving $E_T$ as a free parameter.   

We find that the above equations give results consistent with the exact
diagonalization data, even at rather small spin gap values $E_T\sim t$, as
shown in Fig.~\ref{fig:profiles}(a,c).  Both approaches lead to spectral
functions with a renormalized quasiparticle (qp-) peak whose spectral weight
is transferred to a pronounced hump structure.  [A peculiar momentum
dependence of the matrix elements $\Gamma_{\vc k}$ (note that $\Gamma_{\vc
k=0}=0$) reduces the effect at $\vc k=\Gamma$ point].  Several maxima on the
hump reflect the presence of multiple triplet excitations created by the hole
propagation. All these are the typical signatures of polaron physics.
Multiplet structure of the hump will in reality be smeared by phonons which
are naturally coupled to the $\tilde t$ transition involving also the orbital
sector. Although experiments \cite{Rue06} indicate that electron-phonon
coupling is moderate in cobaltates, it may substantially enhance the
spin-polaron effects, as in case of cuprates \cite{Mis04}. 

\begin{figure}[tbp]
\begin{center}
\includegraphics[width=8.2cm]{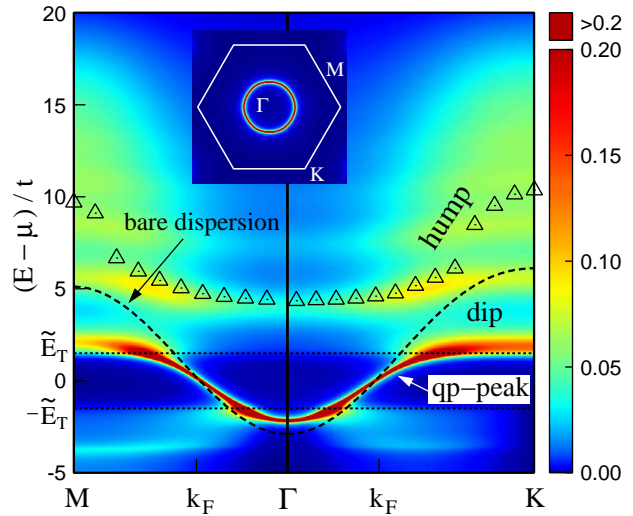}
\caption{
(Color online) Intensity map of the spectral density $A({\vc k},E)$ of the
Co$^{4+}$ holes along M-$\Gamma$-K path in the Brillouin zone calculated at
$30\:\%$ doping and $E_T=2t$.  As the hole energy reaches the renormalized
spin excitation energy $\tilde{E}_T$, the qp-peak broadens and its weight is
transferred to a broad, incoherent structure. This results in a
``peak-dip-hump'' profile of $A({\vc k},E)$ seen also in
Fig.~\ref{fig:profiles}. The top of the smoothed hump structure is indicated
by triangles. The dashed line shows the bare dispersion.  The Fermi surface is
shown in the inset.
}
\label{fig:colormap}
\end{center}
\end{figure}

To illustrate the gross features of the hole renormalization, in
Fig.~\ref{fig:colormap} we show a complete map of the spectral function along
M-$\Gamma$-K path in the Brillouin zone.  We have used a representative value
$E_T=2t$ which is renormalized by holes to $\tilde{E}_T\approx 1.4t$.
Compared to the bare dispersion, the bandwidth of the renormalized holes is
reduced by a factor $\sim 2$. The main observation here is that as the hole
energy reaches $\tilde{E}_T$, the dynamical generation of $S=1$ excitations
becomes very intense and a broad incoherent response develops, leading to the
pronounced ``peak-dip-hump'' structure of $A({\vc k},E)$.  Following the
maximum of the smoothed hump structure, we observe its strong dispersion
(stemming also from incoherent $\tilde t$ hopping). 

\begin{figure}[tbp]
\begin{center}
\includegraphics{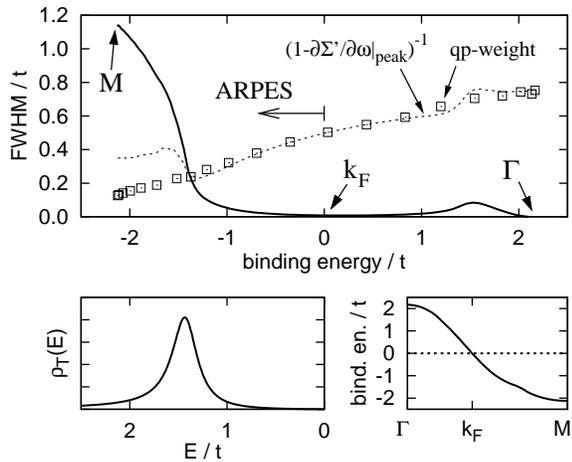}
\caption{
Top panel: Full energy width at half maximum of the quasiparticle peak along
the M-$\Gamma$ dispersion curve (see bottom-right panel) plotted as a function
of the binding energy. The part from $k_F$ to M is accessible by ARPES
experiments.  Strong qp-damping below $\sim -1.4t$ is due to a scattering on
$S=1$ excitations. The spectral weight of the qp-peak obtained by a direct
integration is indicated by squares. When damping is small, it coincides with
a conventional qp-residue $(1-\partial\Sigma'/\partial\omega)^{-1}$.  Lower
panel: The ${\cal T}$-exciton spectral function (left), and renormalized hole
dispersion (right).
}
\label{fig:FWHM}
\end{center}
\end{figure}

Fig.~\ref{fig:colormap} suggests a possible determination of $\tilde{E}_T$
from the quasiparticle damping. To address this problem, in
Fig.~\ref{fig:FWHM} we show the energy width of the qp-peak following its
dispersion curve.  The sharp onset of the damping at the binding energy
$\approx -1.4t$ is clearly related to the maximum of the spin-excitation
spectral function $\rho_T(E)$.  In addition, Fig.~\ref{fig:FWHM} shows the
weight of the qp-peak which is $\vc k$-dependent (mainly due to the matrix
element $\Gamma_{\vc k}$).

Comparison of Figs.~\ref{fig:colormap} and \ref{fig:FWHM} with the data of
Refs.~\onlinecite{Bro07,Qia06} reveals a remarkable correspondence between
theory and experiment. In particular, both the qp-peak and the hump
dispersions (see Figs.~2 and 3 of Ref.~\onlinecite{Bro07}) are well reproduced
by theory, considering $t\approx 100\:\mathrm{meV}$ suggested by the band
structure fit \cite{Zho05}. The onset energy $\tilde{E}_T\sim 1.4t$ for the
qp-damping (Fig.~\ref{fig:FWHM}) is then $\approx 140\:\mathrm{meV}$, in 
nice agreement with experiment (see Fig.~2c of Ref.~\onlinecite{Bro07}).  
Physically, dilute spin-polarons are expected to be pinned by disorder, thus 
qp-peaks should be suppressed at low hole doping. 

To summarize, we have presented a theory for the photoemission experiments in
layered cobaltates. A strong damping of quasiparticles, their reduced spectral
weights, broad and dispersive incoherent structures, and a ``peak-dip-hump''
type lineshapes all find a coherent explanation within our model.  We thus
conclude that unusual correlations observed near the band insulator regime of
cobaltates are the direct manifestation of the spin-state quasidegeneracy of
Co ions, and this intrinsic feature of cobaltates should be the key for
understanding of their unique physical properties, such as the high
thermopower. In particular, the spin-polaronic nature of charge carriers
should be essential for explanation of its remarkable magnetic field
sensitivity \cite{Wan03}.

We would like to thank B.~Keimer for stimulating discussions. This work was
partially supported by the Ministry of Education of CR (MSM0021622410).


\begin{thebibliography}{99}
\bibitem{note1}
$t_{2g}^4e_g^2$ $S=2$ configuration is not accessible by hopping.
\end{thebibliography}
\end{document}